\documentclass{aa}  

\usepackage{graphicx}
\usepackage{txfonts}
\usepackage{lipsum}
\usepackage{subcaption}         
\usepackage{lscape}             
\usepackage{placeins}           
                                
\newcommand{\kilometersecond}{${\rm km}\ {\rm s}^{-1}$}
\newcommand{\metersecond}{${\rm m}\ {\rm s}^{-1}$}

\newcommand{\centimetersecond}{${\rm cm}\ {\rm s}^{-1}$}
\newcommand{\milimitersecond}{${\rm mm}\ {\rm s}^{-1}$}

\newcommand{\derivativemethod}{Derivative}

\usepackage{multirow}
\newcommand{\espresso}{ESPRESSO}

\newcommand{\sbart}{\texttt{s-BART}}

\usepackage{orcidlink}

\begin{document}

   \title{The impact of numerical derivatives on radial velocity extraction}

%

   \author{A. M. Silva\inst{\ref{inst1}, \ref{inst2}} \orcidlink{0000-0003-4920-738X}
          K. Al Moulla\inst{\ref{inst1}}\thanks{SNSF Postdoctoral Fellow}  \orcidlink{0000-0002-3212-5778}
        }
   \institute{
   Instituto de Astrof\'{\i}sica e Ci\^encias do Espa\c{c}o, CAUP, Universidade do Porto, Rua das Estrelas, 4150-762 Porto, Portugal \label{inst1}
   \and Departamento de F\'{\i}sica e Astronomia, Faculdade de Ci\^encias, Universidade do Porto, Rua do Campo Alegre, 4169-007 Porto, Portugal \label{inst2} 
   }
   \date{Received September 30, 20XX}

 
  \abstract
   {The radial velocity (RV) method is a widely used technique to infer planetary masses and orbital parameters. One of the most widely used methods to compute RVs is based on the alignment of a high-SNR, data driven, stellar model with individual observations, commonly referred to as template matching. Typically, the alignment is performed through a $\chi^2$ minimization, but some approaches rely on the derivative of the stellar template to do so, which is often the case in line-by-line methods.}
   {In this paper we aim to explore the limitation of derivative-based methods for RV extraction in the case of using low-SNR stellar models.} 
   {We use simulated Gaussian profiles to investigate the effect of computing a numerical derivative of the stellar template, in comparison with the usage of an analytical profile. The impact on RV and associated uncertainty is then analyzed as a function of the signal-to-noise ratio (SNR) of the line. Then, using real observations we compare the residuals between a derivative-based RV extraction and a classical template matching implementation, as function of the SNR of the stellar template.}
   {We find that on simulated Gaussian profiles the usage of the numerical approach leads to a RV residual at the level of the meter per second, at a per-pixel SNR regime of 100. An increase of the SNR leads to a decrease of this impact, falling under the current noise-floor of state-of-the-art spectrographs at SNR>1000 for a single spectral line. The inclusion of multiple spectral lines in the simulations lead to an overall decrease of the contamination, across all SNR regimes. The application to a real dataset presents a decrease of the RV impact with the increase of the template's SNR, albeit still presenting a 12.5 \centimetersecond{} difference when including 78 observations in the stellar template.}
   {}


   \maketitle
\nolinenumbers

\section{Introduction}

   Since the discovery of the first exoplanet orbiting a Sun-like star in 1995 \citep{mayorJupitermassCompanionSolartype1995}, more than 6000 exoplanets have already been detected\footnote{Information from the 15th January, 2026. Refer to \url{https://exoplanetarchive.ipac.caltech.edu/} for updated  information.}. However, one of the boldest goals of modern-day astrophysics still stands: the detection and characterization of an Earth-twin - a rocky world, orbiting a solar-like star, that has the physical conditions to hold liquid water in its surface. One of the major workhorses of the field lies in the radial velocity (RV) method, based on measuring variations on the line-of-sight velocity of a star. 

   The extraction of RVs from spectra is typicallly  accomplished through a cross-correlation of a weighted mask with the spectra \citep{baranneELODIESpectrographAccurate1996,pepeCORALIESurveySouthern2002}. However, this technique is not always the optimal approach, especially in the cooler M-dwarfs where the template matching (TM) algorithm \citep{anglada_escude_HARPS_TERRA_2012,zechmeisterSpectrumRadialVelocity2018,silvaNovelFrameworkSemiBayesian2022} usually lead to reduced RV scatter and improved precision. In this technique, a data-driven stellar template is constructed by stacking multiple observations of the same star, and then aligned with the individual observations. More recently, the application of line-by-line (LBL) techniques  allows the rejection of highly variable lines through different outlier rejection criteria. LBL RVs can be extracted either through TM applied to small wavelength intervals \citep{dumusqueMeasuringPreciseRadial2018, artigauLinebylineVelocityMeasurements2022,almoullaARVEAnalyzingRadial2025} or by directly modelling stellar lines with e.g. Gaussian profiles \citep{lafargaCARMENESSearchExoplanets2020a,sannicolasmartinezTILARATemplateIndependentLinebyline2026}.

   The determination of precise RV measurements faces significant challenges and contamination from spurious signals, with amplitudes often surpassing those of the signature of an Earth-like planet orbiting a Sun-like star ($\sim$ 9 \centimetersecond). The largest contributor to this lies on the host star, introducing signals with varying impacts both over wavelength and time \citep{oshaghUnderstandingStellarActivityinduced2017,costesLongtermStellarActivity2021, almoullaMeasuringPreciseRadial2022}. To overcome such efforts, we have seen the advent of multiple solar telescopes, observing the Sun as if it was a different star and allowing a translation of insights to analyze data from others stars \citep[e.g., ][]{dumusqueThreeYearsHARPSN2021,linObservingSunStar2022, farretjentinkABORASPolarimetric10cm2022, almoullaStellarSignalComponents2023, rubenzahlStaringSunKeck2023,llamaLowellObservatorySolar2024,santosPoETParanalSolar2025}, albeit not yet reaching the 10 \centimetersecond{} frontier.
   
   More recently, it has been shown that there are other effects conditioning our current ability to detect and characterize small RV signals. More specifically, the RV extraction methods that rely on a data-driven template can be affected by systematics intrinsic to the method:

   \begin{enumerate}
      \item \citet{silvaSystematicBiasTemplatebased2025} has show that in cases where the BERV separation between observations is small, micro-telluric and other detector-frame systematics can introduce multi-meter-per-second quasi-linear trends on the RVs;
      \item \citet{doshiInterpolationConstraintRV2025} has shown, through simulations, that the interpolation of stellar spectra can introduce biases in the data at a sub-\metersecond level, whilst \citet{silva_impact_nodate} revealed that in real observations such effects are measurable and can surpass the \metersecond{} barrier under specific conditions.
   \end{enumerate}

   In this paper we present a new source of systematic errors in RV extraction, present in methods that rely on the numerical computation of the derivative of the stellar spectra \citep{connesAbsoluteAstronomicalAccelerometry1985,bouchyFundamentalPhotonNoise2001}, which is often the case for LBL approaches. In Section \ref{Sec:methods} we describe the formalism for RV extraction that will be used to characterize such effects. Section \ref{Sec:gauss_profiles} we evaluate the systematic effect with Guassian profiles, following by an application to real observations in Section \ref{Sec:ESPRESSO_result}. Lastly, in Section \ref{Sec:conclusions} we discuss the findings of this paper and contextualize them in the current state of the field.

\section{Methods} \label{Sec:methods}

   The extraction of RVs through TM methods typically follow the same underlying principles:

   \begin{itemize}
      \item We start by constructing a high signal-to-noise ratio (SNR) stellar template, that is taken to be a time-invariant model capable of perfectly describing each observation of a given star. 
      \item At the level of each spectral order (or line, for LBL applications) we find the optimal RV displacement that bests aligns the observed data and model.
      \item The individual RV measurements are combined through a weighted mean, or equivalent, creating a single RV measurement for a given observation. 
   \end{itemize}

   In this paper we will apply two formulations of the template matching algorithm, extracting RVs at the order of each spectral order. This will be done under two theoretically equivalent paradigms: i) a classical template TM formalism; ii) a methodology based on a Taylor expansion of the stellar template \citep{connesAbsoluteAstronomicalAccelerometry1985,bouchyFundamentalPhotonNoise2001}, that will be referred to as "\derivativemethod" method. The following subsections will describe the steps in the RV extraction process.

   \subsection{Classical template matching} \label{SubSec:classicalTM}

      In a classical TM framework the template is aligned with the individual observations through a $\chi^2$ minimization procedure applied at the level of each spectral segment (e.g., individual orders or spectral lines):

      \begin{equation}
         \chi^2 = \sum_{i=1}^{N_{pixels}} \frac{1}{\sigma_S^2 + \sigma_T^2 } \cdot [S_{\lambda_i} - \mathcal{P}_{\lambda_i} \cdot T_{\lambda_i}]^2
      \end{equation}
      
      \noindent where $S_{\lambda_i}$ is the flux of the stellar spectrum, $\sigma_S$ its associated uncertainty, $\mathcal{P}_{\lambda_i}$ a first degree polynomial whose parameters are fitted to account for changes in the continuum level in relation to the stellar template, $T_{\lambda_i}$ and $\sigma_T$ the stellar template and associated uncertainty, all evaluated at wavelength $\lambda_i$.
      
      The spectral segments are assumed to be independent measurements of the same underlying variable, and as such the piece-wise RVs are combined through a weighted average:

      \begin{equation}
         RV = \frac{\sum RV_i \cdot \sigma_i^{-2}}{\sum \sigma_i^{-2}}
      \end{equation}

      \noindent where $RV_i$ is the RV measured in the \textit{i}-th spectral order, and $\sigma_i$ the associated uncertainty. In order to avoid inconsistencies in the extracted RVs, we also ensure that all observations use exactly the same spectral orders, as the pre-processing stage might reject poor quality ones. As such, if any order is rejected in any given observation, it will also be rejected in all others. The extraction of RVs through this classical formalism is done through the \sbart{} \citep{silvaNovelFrameworkSemiBayesian2022} package\footnote{Publicly available at \url{https://github.com/iastro-pt/sBART}}. We refer the reader to the \sbart{} manuscript and documentation for further details on the implementation.

   \subsection{Taylor-expansion based template matching} \label{Subsec:TaylorTM}

      An alternative approach to $\chi^2$ minimization for RV extraction (as in Sect. \ref{SubSec:classicalTM}) is to describe the RV displacement as 

      \begin{equation} \label{Eq:taylor_formula}
         \frac{\delta RV}{c} = \frac{S(\lambda) - T(\lambda)}{\lambda \cdot \partial T / \partial \lambda}
      \end{equation}
      
      \noindent where $\partial T / \partial \lambda$ is the derivative of the stellar template, as a function of wavelength. Similarly, the associated RV uncertainty per pixel can be obtained through

      \begin{equation} \label{Eq:taylor_formula_per_pix}
         \sigma_{v, pix}\approx\left|\frac{\partial\ T}{\partial\ \lambda}\right|_i^{-1} \frac{c}{\lambda_{pix}} \sigma_{S, pix}
      \end{equation}
      \noindent and converted to a final RV uncertainty per spectral region as
      
      \begin{equation}\label{Eq:taylor_formula_err}
         $\[\sigma_{\mathrm{RV}}=\left(\sqrt{\sum_i \frac{1}{\sigma_{v, {pix}}^2}}\right)^{-1}.\]$
      \end{equation}

      This approach makes use of a first-degree Taylor expansion, coupled with the assumption of a noise-free stellar template. It was proposed by \citet{connesAbsoluteAstronomicalAccelerometry1985} and is implemented within multiple state-of-the-art RV extraction pipelines. When using the same stellar template and the same dataset, the RVs determined through Eq. \ref{Eq:taylor_formula} should be equivalent to those that are derived through a traditional $\chi^2$ minimization.
      
      For our application, \sbart{} package was adapted to also allow RV estimation through Eq. \ref{Eq:taylor_formula} and Eq. \ref{Eq:taylor_formula_err}. The derivative of the template, $\partial\ T/\partial\ \lambda$,  is computed through a second order accurate central differences approach, as implemented in \textit{NumPy} \citep{harrisArrayProgrammingNumPy2020}. Similarly to the approach of our classical TM, we leverage a first degree polynomial to match the continuum level of the template and the one of the individual observations. After computing an RV estimate for all valid spectral segments, the measurements are combined through the weighted mean approach of Sect. \ref{SubSec:classicalTM}. 

\section{Application to Gaussian profiles} \label{Sec:gauss_profiles}

   In order to evaluate if the "\derivativemethod{}" formalism for RV extraction (Sect. \ref{Subsec:TaylorTM}) is affected by possible numerical errors in the computation of the derivative of the stellar template, we start by analyzing two simplistic cases: i) a single Gaussian profile with injected noise, and ii) a sum of N Gaussian profiles.The Gaussian profile is given by:
   
   \begin{equation} \label{Eq:gauss_line}
      \mathcal{F(\lambda)} = 1\ -\ a\ \cdot\ e^{-\frac{(\lambda - \lambda_{c})^{2}}{2\ \cdot\ \sigma^2}}
   \end{equation}

   \noindent where \textit{a} represents the normalized line depth, $\lambda_c$ the central wavelength of the line, $\sigma$ its standard deviation (i.e., the line width), and $\lambda$ the wavelengths for which the line is defined.  The wavelength solution of this Gaussian profile is defined to be uniformly separated which, as a first order approximation, holds true in real spectrographs within a small wavelength interval. The $\Delta\lambda$ is given by $\sigma\ /\ N_{pix}$, where $N_{pix}$ is taken to be the number of pixels per line width, a metric of how well sampled the spectral line is. Lastly, we inject noise into the spectra through a Poisson distribution, so that a given SNR is reached. For this analysis, the stellar template is also built from Eq. \ref{Eq:gauss_line}, with an SNR defined to be 10 times larger than the one in the observation

   The RV extraction, through the "\derivativemethod" method, will be tested when using both the numerical derivative of the template (through finite differences) and through its analytical formulation (Eq. \ref{Eq:analytic_guass_deriv}):

   \begin{equation} \label{Eq:analytic_guass_deriv}
      \frac{\partial \mathcal{F}(\lambda)}{\partial \lambda}  = \frac{(\lambda\ -\ \lambda_{c})}{\sigma^2}\ \cdot a\ \cdot\ e^{-\frac{(\lambda - \lambda_{c})^{2}}{2\ \cdot\ \sigma^2}}
   \end{equation}

   \subsection{Single Gaussian}
      The impact on RV extraction, at the level of a single Gaussian line, will be evaluated through different line profiles, with the following possible parameters:
         
         \begin{itemize}
            \item Normalized depths that can take any of the following values $[0.15,\ 0.5,\ 0.75]$;
            \item Widths of the Gaussian lines of [6, 8, 10] \kilometersecond;
            \item Number of pixels per line width of [5, 10, 15, 20];
            \item Signal-to-Noise ratio (SNR) ranging from $10^2\ to\ 10^6$.
         \end{itemize}
         
\begin{figure}[h!]
            \centering
            \resizebox{\hsize}{!}{\includegraphics{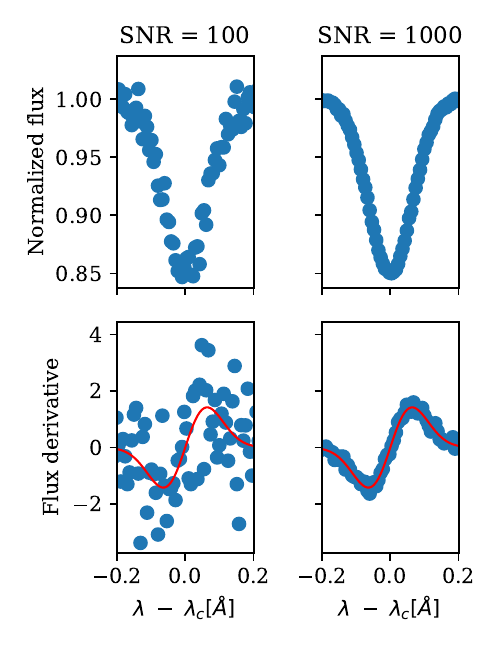}}
            \caption{\textbf{Top panel}: Representation of a Gaussian with Poisson noise added into it for two SNR values. \textbf{Bottom panel}: Comparison of the numerical derivative of the profile (blue) with the analytical profile from Eq. \ref{Eq:analytic_guass_deriv} (red).}
            \label{Fig:gaussian_deriv_comp}
         \end{figure}

         From every combination of the aforementioned parameters we construct 2000 Gaussian profiles, through to independent realizations of the Poisson noise. In Figure \ref{Fig:gaussian_deriv_comp} we show one Gaussian profile and the comparison of its numerical and analytical derivatives.
         
\begin{figure*}[ht]
            \centering
               \includegraphics[width=17cm]{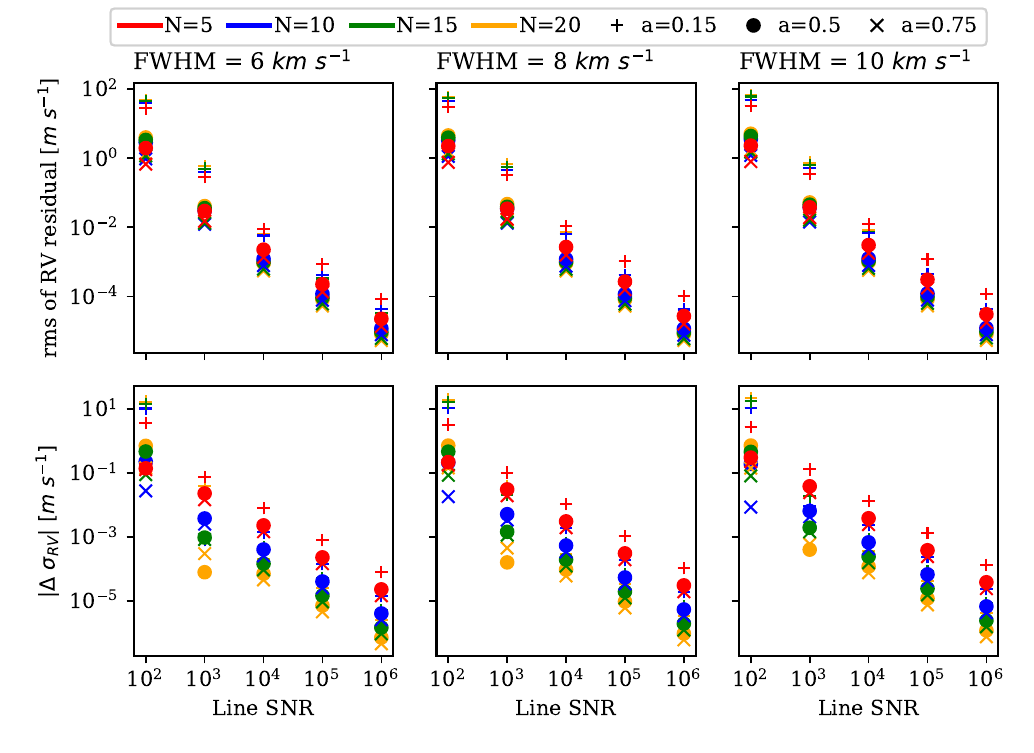}
               \caption{Comparison of RVs extracted with the numerical and analytical derivation schemes from 2000 Gaussian Profiles with injected Poisson noise. The colors represent the number of pixels per FWHM of the line, whilst the markers provide the normalized line depth for the simulations. \textbf{Top panel}: Standard deviation of the RV residuals as a function of SNR of the simulated Gaussian, when using a stellar template with an SNR 10 times larger; \textbf{Bottom} panel: Median value of the absolute difference in RV uncertainty provided by the two methods of deriving the stellar template.}
               \label{Fig:gaussian_noise_rv_comp}
         \end{figure*}

         As one would expect, we find that the injected flux noise leads to a contamination of the first derivative of the data, distorting the true profile that one would expect for the derivative (bottom panel). It follows that such distortion would be propagated to the RV extractions stage, biasing the measurement. Figure \ref{Fig:gaussian_noise_rv_comp} shows the impact of applying a numerical derivation technique instead of the analytical, focusing on the RV scatter and the associated uncertainties.

         Through this analysis we find that the impact on RV and uncertainty is clearly correlated with the SNR of the simulations, showing larger differences between the numerical and analytical computation for the noisier data. This is to be expected, as the noisier data will lead to sharper discontinuities and the spectra and, as such, lead to deviations from the true derivative profile. Overall, the sampling of the lines does not play a large impact, as the results are consistent for any of its three possible values. The driving factor for the magnitude of the impact is the line depth, with shallower lines leading to a typically larger scatter of the RV residuals and leading to larger differences in the estimate RV uncertainty. It is also important to note that the reported RV impact  (between using a numerical and analytical derivations) fall to levels below the expected precision of state-of-the-art spectrographs for a per-pixel SNR larger than $10^3$. For the estimation of the RV uncertainty we find that the wrong determination of the template's gradient does not play a relevant role for per-pixel SNR larger than $10^2$.
      
   \subsection{A forest of Gaussian profiles}
         
      To evaluate the dependency of the RV scatter with the number of lines we fix the line properties, to a FWHM of 10 \kilometersecond{} and a normalized depth of 0.5. For this analysis, we construct the line centers in such a way that there are no blended lines and inject Poisson noise as per our previous formulation. In this analysis, we also vary the SNR of the template, exploring different regimes of application of the TM technique, which are typically found in their application to real datasets.

      \begin{figure}[h!]
         \centering
         \resizebox{\hsize}{!}{\includegraphics{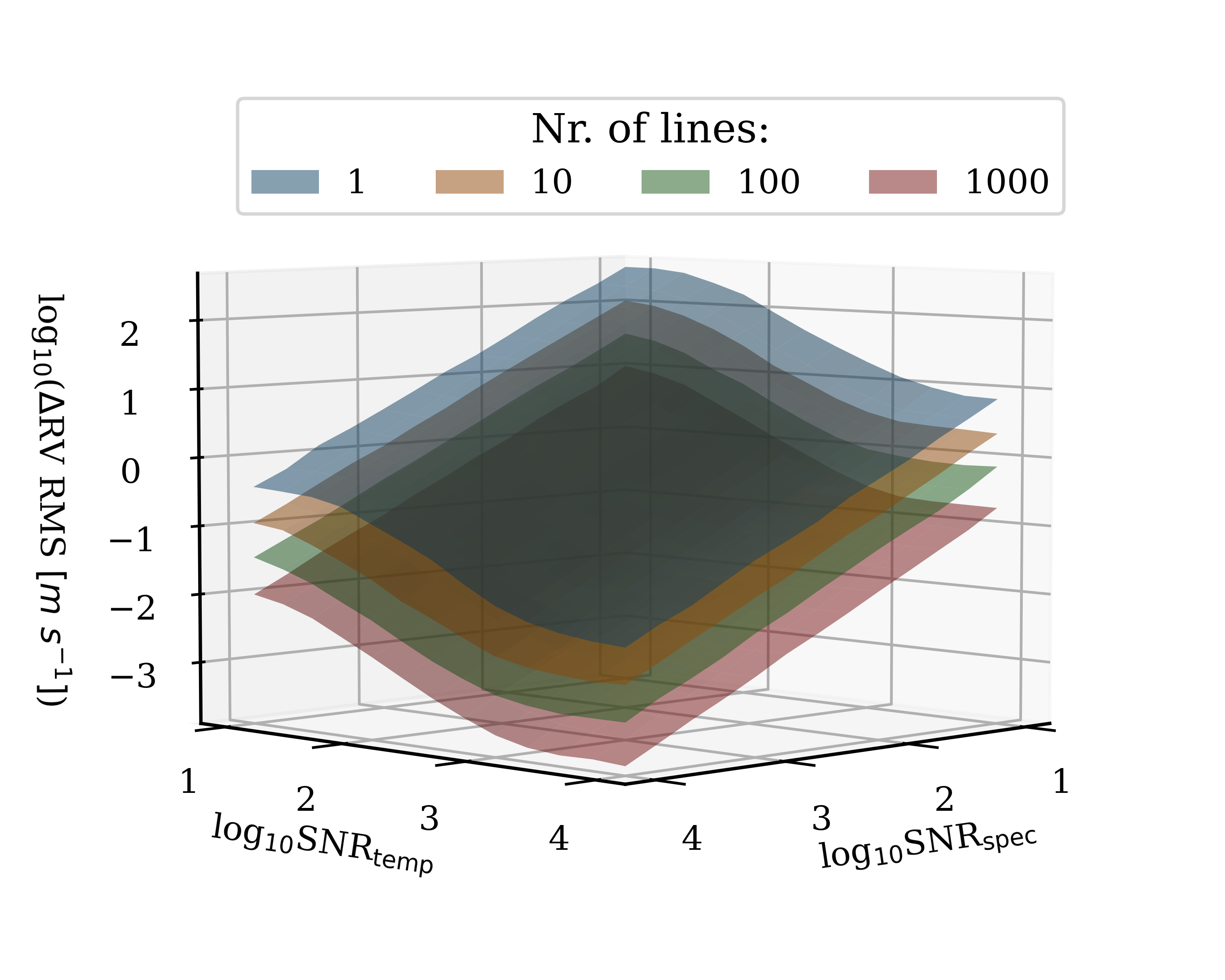}}
         \caption{Comparison of RVs extracted with the numerical and analytical derivation schemes, with spectra that contains different number of lines. Different SNR ratios between template and spectra are explored.}
         \label{Fig:Ngaussian_noise_rv_comp}
      \end{figure}

      Figure \ref{Fig:Ngaussian_noise_rv_comp} presents the result of this comparison, showing a decrease of the difference between the two formulations and the number of lines increase (from a single line in blue, to 1000 lines in red). Furthermore,  we find that an increase of the SNR, of either the template or the data, leads to an overall decrease of the RV difference. This agrees with our previous conclusions, as the lower noise level leads to smaller inaccuracies on the numerical derivation of the flux. 

\section{Application to ESPRESSO observations} \label{Sec:ESPRESSO_result}

   In Section \ref{Sec:gauss_profiles} we found that the calculation of a numerical derivative introduces measurable effects on Gaussian profiles in lower-SNR regimes. The same exercise cannot be repeated with real observations as we do not know the true derivative of the stellar spectra. We can, however, compare the RVs extracted through the "\derivativemethod{}" and the classical methods. In order to do so, we use the \espresso{} observations of K2-129 sequentially selecting the observations that are used in the template. In total, we have 78 observations of this M3 star \footnote{Observations collected at the European Southern Observatory under ESO program IDs: 1104.C-0350(I), 1104.C-0350(V), 108.2254.006, 1104.C-0350(S), 1104.C-0350(X), 1102.C-0744(N), 108.2254.005, 1104.C-0350(P), 1104.C-0350(M), 1104.C-0350(L), 1104.C-0350(Q), 1102.C-0744(Z), 1104.C-0350(T), 110.24CD.003, 1104.C-0350(B), 1102.C-0744(M), 106.21M2.002, 1104.C-0350(R), 106.21M2.007, 1102.C-0744(V)} collected between Jul 20, 2019 and March 25, 2023. The spectra has a median flux SNR of 25 on \espresso's order 120, which is centered at $\sim$ 4571 $\AA$, and a median CCF RV uncertainty of $\sim$ 1 \metersecond. This star was selected due to it's relatively low SNR of individual observations, allowing us to probe the impact of the numerical derivative on the RV extraction in the noisier regime that was identified with the simulations of Sect. \ref{Sec:gauss_profiles}.
   
   Our objective is to bin our data such that we can generate an increase of the template's SNR, which we accomplish through a sequential addition of observations to the template. For this purpose, we apply the following recipe:

   \begin{itemize}
      \item We select the first \textit{N} observations, from which we will construct the stellar template;
      \item Using the stellar template that was constructed in 1) we compute RVs through the two methods presented in Sect. \ref{Sec:methods}. We shall refer to those of Sect. \ref{SubSec:classicalTM} as  "classical" and the ones from Sect. \ref{Subsec:TaylorTM} as "\derivativemethod" method;
      \item We increase \textit{N} and repeat from step 1) until all observations are used for the construction of the template.
   \end{itemize}
   
\begin{figure}[h!]
      \centering
      \resizebox{\hsize}{!}{\includegraphics{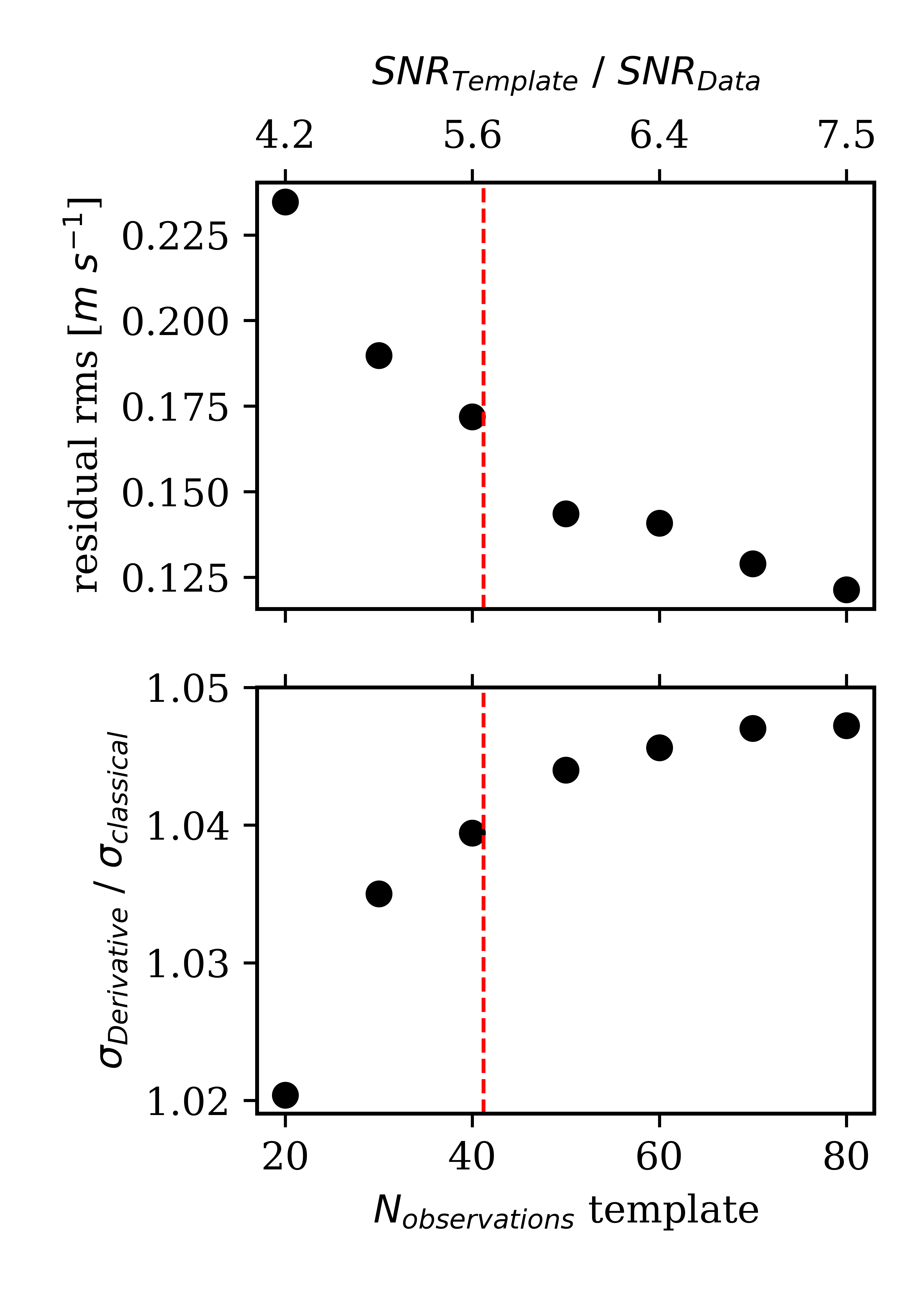}}
      \caption{Evolution of the difference between the \derivativemethod{} and the classical method, as a function of the number of observations used in the construction of the template (bottom axis) and the ratio of the SNR of the template and the median SNR of the observations in \espresso{} order 120 (top axis). The top panel presents the evolution of the standard deviation of the RV residuals, whilst the bottom one presents the ratio of the median RV uncertainty. The vertical red line represents the 80th percentile of the number of observations in \espresso's data release \citep{segransanESPRESSOLegacyData2025}.}
      \label{Fig:k2_129_snr_evol}
   \end{figure}

   We refer the reader to Appendix \ref{App:data_pre_process} for details on data pre-processing  that is applied to \espresso{} observations to ensure quality spectra. Figure \ref{Fig:k2_129_snr_evol} presents the results of such comparison, in agreement with the results of the previous Section. 

   We find that as the SNR of the template increases the RVs yielded by the two measurements start agreeing with one another, albeit with a scatter above the 10 \centimetersecond{} level. In the lower SNR regime, with close to 20 observations being used for the construction of the template, we find an impact almost two times larger. Interestingly, the same result is not recovered for the RV uncertainties, as the SNR increase  translates into an increase in the median RV uncertainty, albeit still in agreement with the one recovered with the one from the classical method. This agreement is expected, as the two approaches are using exactly the same stellar template to derive RVs and associated uncertainties. 

   As observation time is not unlimited, real observation campaigns stop whenever the main science goal has been achieved which, depending on the star's SNR, might not allow the construction of a sufficiently high-quality stellar template. An example of such lies in \espresso's data release \citep{segransanESPRESSOLegacyData2025}, where the 80th percentile of stars was observed $\sim$ 41.2 times. It is possible that the RV extraction through approaches that rely on the numerical computation of the derivative could be biased by numerical errors, with residual scatter comparable to the signature of an Earth-twin, posing challenges for the detection and characterization of low-mass exoplanets.

\section{Conclusions and Discussion} \label{Sec:conclusions}

   In this paper we present an analysis of one of the core assumptions behind one commonly used framework for the extraction of precise RVs. The reliance on the computation of a numerical derivative of the stellar template, often assumed to be noise free, leads to the inclusion of numerical errors into the determination of the optimal RV for the alignment of observations and model. In particular, we show that this is the case using purely Gaussian profiles with injected noise, injected through a Poisson distribution. A comparison of the RVs yielded when applying a numerical derivative to those obtained when using its analytical profile reveals a clear correlation with the SNR of the data. We find that on a single line, with a per-pixel SNR of 100 we retrieve RV residuals at the order of the 1 \metersecond or higher, depending on line depth. The increase of the SNR leads to its decrease, until a point in which the  RV residual lies below the noise floor of current instruments. Not only do we find an impact on the RVs, but also on the associated RV uncertainties as they also include information from the derivative of the data. It is however important to note that the residual of RV uncertainties presents the same correlation with SNR, with the effect reaching the \milimitersecond{} for per-pixel SNR's greater than 100. Furthermore, an increase of the number of lines in the spectra also translates into a decrease of the contamination floor, across all tested SNR regimes.

   Our analysis of K2-129 \espresso{} observations reveal that the same behavior is recovered on real data, as we increase the SNR of the template. By sequentially adding more observations on the stellar template, we find that the RVs yielded by a "\derivativemethod{}" method and a classical template matching algorithm converge towards agreement, albeit still presenting a 12.5 \centimetersecond{} residual signal with 78 observations in the template, reaching a template SNR 7.5 times greater than that in the individual observations. Furthermore, a comparison of the RV uncertainties also shows an increase ratio of RV uncertainty over time, albeit it is not straightforward to draw conclusions from this result as the two methods estimate the RV uncertainty through different avenues.

   The estimation of spectral derivatives through numerical methods, in the presence of noise, has shown a significant impact in RV extraction, with the possibility to also extend to other science cases that make use of such approaches. Not only does it impact radial velocity scatter, but noisy derivatives could also artificially inflate the RV information, leading to a possible underestimation of RV uncertainties. We posit that some mitigation strategies could entail a smoothing of either the spectra or its numerical derivative, decreasing the impact of high-frequency flux variation on the per-pixel velocity estimate.

\begin{acknowledgements}
   The authors thank E. Artigau and N. Cook for fruitful discussions. This work was funded by the European Union (ERC, FIERCE, 101052347). Views and opinions expressed are however those of the author(s) only and do not necessarily reflect those of the European Union or the European Research Council. Neither the European Union nor the granting authority can be held responsible for them. This work was supported by Fundação para a Ciência e a Tecnologia (FCT) through national funds under the research grant UID/04434/2025 (DOI 10.54499/UID/04434/2025). KA acknowledges support from the Swiss National Science Foundation (SNSF) under the Postdoc Mobility grant P500PT\_230225.
\end{acknowledgements}

\bibliographystyle{aa} 
\bibliography{derivative_paper} 

\begin{appendix}
   \nolinenumbers

   \section{Data pre-processing} \label{App:data_pre_process}

      The extraction of precise RVs is based upon the usage of high-quality stellar spectra, raising the need of pre-processing the observations to reject bad pixels and other sources of contamination in the stellar spectra. We reject all pixels that were flagged by \espresso's Data Reduction Pipeline (DRP), encompassing hot pixels, cosmic rays, and other pixels that do not meet the quality standards. Then, we construct a binary mask to reject any wavelengths in which there are telluric lines with depths larger than 1\%. This binary mask is constructed from a synthetic spectrum of Earth's transmittance, built with \texttt{TelFit} \citep{gulliksonCorrectingTelluricAbsorption2014} using the atmospheric conditions of the night with the highest humidity content. The mask also accounts for every possible BERV of the observation, thus considering the different positions of the atmospheric imprint on the data.

      The last step in the pre-processing stage is the construction of the stellar template, which is done after placing each observation in a common referential, as given by previous RV measurements of the observations. After that point, the observations are interpolated with a cubic spline algorithm to a common wavelength grid, and then stacked through a weighted mean. 

      In this paper we use the \texttt{ASTRA} \citep{silvaASTRAPythonPackage2026} package\footnote{Publicly available at \url{https://github.com/Kamuish/ASTRA}} to interface with the observations and construct the stellar and telluric templates. During its application, spectral orders can be fully rejected if they do not meet the pre-defined quality criteria. We refer the reader to \texttt{ASTRA}'s documentation for further details.

\end{appendix}
\end{document}